\documentclass[a4paper]{jpconf}
\usepackage{graphicx}
\usepackage{amsmath}
\usepackage{amssymb}

\begin{document}

\title{Mass transfer during drying of colloidal film beneath a patterned mask that contains a hexagonal array of holes}
\author{Yu Yu Tarasevich, I V Vodolazskaya}
\address{Astrakhan State University, Russia}
\ead{tarasevich@asu.edu.ru}

\begin{abstract}
We simulated an experiment in which a thin colloidal sessile droplet is allowed to dry out on a horizontal hydrophilic surface when a mask just above the droplet predominantly allows evaporation from the droplet free surface directly beneath the holes  in the mask [Harris D J, Hu H, Conrad J C and Lewis J A 2007 \textit{Phys. Rev. Lett.} \textbf{98} 148301]. We considered one particular case when centre-to-centre spacing between the holes is much less than the drop diameter. In our model, advection, diffusion, and sedimentation were taken into account. FlexPDE was utilized to solve an advection-diffusion equation using the finite element method. The simulation demonstrated that the colloidal particles accumulate below the holes as the solvent evaporates. Diffusion can reduce this accumulation.
\end{abstract}

\section{Introduction}\label{sec:intro}
In our research, we try to simulate the results of the actual experiments~\cite{Harris2007PRL,Harris2008Langmuir,Harris2009PTRS}. When a colloidal film dries under a mask, pattern formation can be observed. The mask induces periodic variations between regions of free and of hindered evaporation. The inhomogeneous evaporation induces flows in the film. The flows carry the colloidal particles towards regions of higher evaporative flux. The particles therefore accumulate under the holes in the mask. Completely dried films exhibit patterns. These patterns can be regulated by the mask geometry, the distance between the mask and the underlying film, and the initial concentration of the colloidal particles. In fact, this pattern formation is a kind of well-known coffee-ring effect~\cite{Deegan1997}. Our model is based on the use of an advection-diffusion equation~\cite{Widjaja2008,Okuzono2009,Maki2011,Tarasevich2013,Eales2015}.

\section{Model}\label{sec:model}
In the experiments~\cite{Harris2007PRL,Harris2008Langmuir,Harris2009PTRS},
the diameter of the droplet, $d$, was about 15~mm and initial height, $h_0$, was about 100~$\mu$m, i.e. $h_0/d \simeq 0.01$. The patterned mask contained a hexagonal array of holes with diameter $d_h = 250 \mu$m, i.e. $d_h/d \simeq 0.01$. Centre-to-centre spacing between the holes, $P$, varied from $2d_h$ to $10d_h$, i.e. $P/d \simeq 0.1$. As a first approximation, the film and the mask can be treated as infinitely large. In this case, only one hexagonal cell is needed for consideration due to translational  symmetry. The behaviour of the such cell is expected to be similar with a drop covered by a lid that had only a small hole over the centre of the drop through which the vapor could escape. The lid restricted the evaporation from the perimeter.  The resulting deposit was uniform rather than being concentrated at the edge~\cite{Deegan2000}. This situation has been simulated using lubrication approximation~\cite{Fischer2002}. Recently, effect of diffusion and hole radius on the component redistribution inside the droplet has been investigated numerically~\cite{Vodolazskaya2014BPFUR}. The opposite limiting case, when the distance between the holes is comparable with the diameter of the drop ($P \approx d$) and the drop itself can be treated as a spherical cup, has been considered~\cite{tarasevich2015EPJE}.

\begin{figure}[ht]
 \centering
\includegraphics[width=0.75\linewidth]{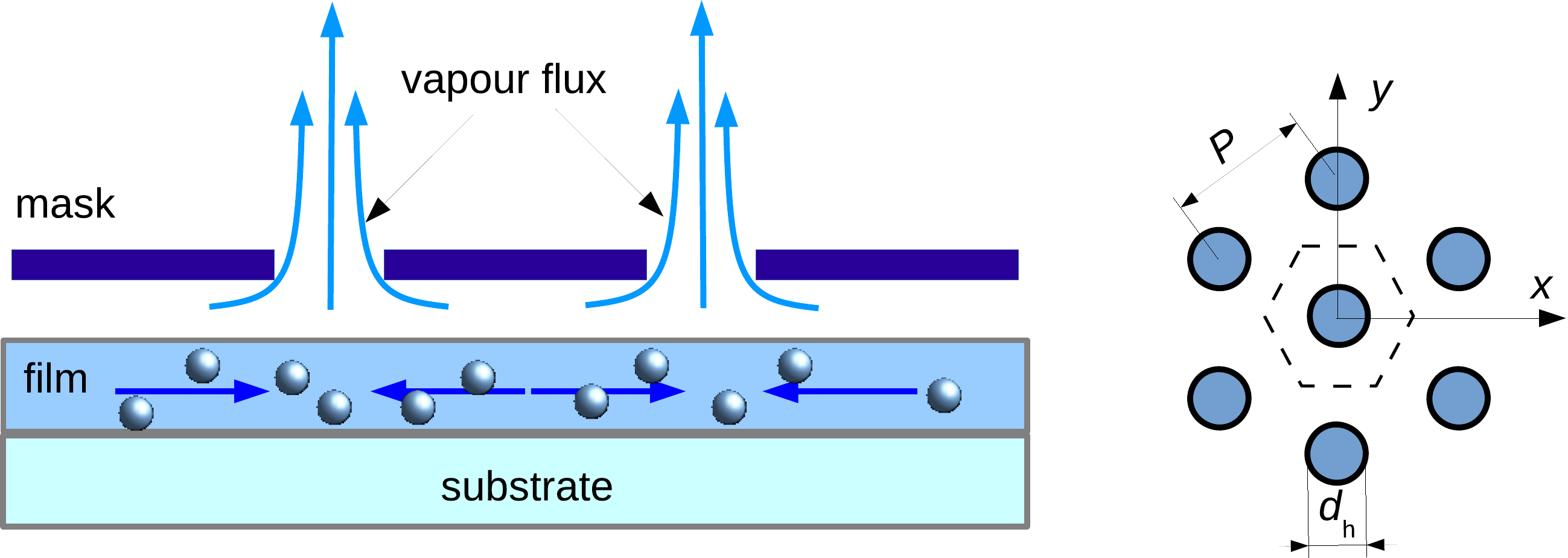}
  \caption{Sketches of a colloidal film beneath a mask (left) and hexagonal array of holes in the mask (right)}\label{fig:film}
\end{figure}

We considered a thin colloidal film on a  horizontal substrate (Fig.~\ref{fig:film}). The substrate was assumed to be solid, hydrophilic, impermeable, and indissoluble by the solvent. We supposed that the suspended particles have no effect on the hydrodynamics. Moreover, we assumed that volume fraction of the colloidal particles, $\Phi$, is small in any internal volume of the film at any time ($\Phi \ll 1$). For the coffee-ring effect, a model accounting finite volume  of the solute particles in evaporating sessile drops of a colloidal solution on a plane substrate was proposed in~\cite{Popov2005}.

Under the normal ambient conditions (room temperature, normal atmospheric pressure, room humidity, unheated substrate), evaporation of the solvent can be considered as slow and the flows inside the droplets can be treated as steady state~\cite{Popov2005}. We ignored any thermal effects and suppose that the surface tension is independent of the particle concentration, thus Marangoni convection cannot occur. We assumed that the diffusivity is constant. Within continual approaches, any effects connected with the particles are ignored. Among these effects are jamming~\cite{Marin2011PRL,Shao2014PhysB} and shape-dependent capillary interactions~\cite{Yunker2011}.

We assumed that the liquid is incompressible. Mass conservation yields
$\nabla \mathbf{v} = 0.$
The boundary conditions are
\begin{itemize}
  \item at the substrate ($z=0$)
  $\partial_z u =0$, where $\mathbf{v} =- \nabla u$,
\item at the imaginary walls of the hexagonal cell
 $
\partial_\mathbf{n} u  =0$, where $\mathbf{n}$ is the outward normal unit vector at the wall,
  \item at the free surface ($\Sigma$) $-\partial_\mathbf{n} u= \mathbf{v}_s \mathbf{n} + \frac{J}{\rho}$ or  $(\mathbf{v} - \mathbf{v}_s)\mathbf{n} = \frac{J}{\rho}$, where  $\mathbf{n}$ is the outward normal unit vector at the liquid-air interface, $J$ is the density of the vapor flux above the free surface and $\rho$ is the density of liquid.
\end{itemize}

Mass transfer inside the film can be described using the advection--diffusion equation~\cite{Widjaja2008,Maki2011}
\begin{equation}\label{eq:advdiff}
  \partial_t c + \mathbf{v} \nabla c =  D\nabla^2 c,
\end{equation}
where $c$ is the concentration of the suspended particles, $D$ is their diffusivity, $\mathbf{v}$ is the velocity of the flow inside the film. Eq.~\eqref{eq:advdiff} should be solved in the region with a moving boundary, because the free surface of the film goes down due to evaporation.

The advection--diffusion equation~\eqref{eq:advdiff} was solved subject to the boundary conditions
\begin{itemize}
  \item at the free surface ($\Sigma$) $c \mathbf{v}\mathbf{n} - D \partial_{\mathbf{n}} c= c \mathbf{v}_s\mathbf{n}$,
where $\mathbf{n}$ is the outward normal unit vector at the liquid-air interface, $\mathbf{v}_s$ is the velocity of the free surface,
  \item at the substrate ($z=0$)
$\partial_z c= -\tilde{k}_d / c$, where $\tilde{k}_d$ is the deposition rate constant~\cite{Widjaja2008}.
\item at the imaginary walls of the hexagonal cell
 $c \mathbf{v}\mathbf{n} - D \partial_{\mathbf{n}} c = 0$, where $\mathbf{n}$ is the outward normal unit vector at the wall.
\end{itemize}

To mimic the influence of a mask above the film, the speculative evaporative flux function was utilized
\begin{equation}\label{eq:J}
J(x,y) = \begin{cases}
      J_0 , & \text{if } \sqrt{x^2 + y^2} < r, \\
      0, & \text{otherwise},
    \end{cases}
\end{equation}
where $x,y$ are the coordinates, $J_0$ is the constant, and $r = d_h/2$. This formula qualitatively correctly reproduces the features of the evaporative landscape above the drying films calculated using finite-element modeling~\cite{Harris2007PRL}.

The free surface of the film was assumed to be flat. Due to evaporation, the film height, $h=h(t)$, changes with time as
\begin{equation*}\label{eq:dVdt}
  \frac{dh}{dt} = -\frac{1}{\rho A_6}\int_{A_6} J(x,y)\, d A = -\frac{a J_0 }{\rho},
\end{equation*}
where $a = \frac{\pi r^2}{A_6}$ is the ratio of a hole area to the area of the base of a hexagonal cell $A_6=2\sqrt{3} (P/2)^2$.

All coordinates were scaled by the $P/2$. The time taken for the complete evaporation of the film, $t_0 = \frac{\rho h(0)}{a J_0}$, is the natural characteristic time.
The characteristic velocity is
$v_0 = \frac{a J_0 }{\rho}.$
We defined the dimensionless time as
$\tau = \frac{t}{t_0}= \frac{a J_0 t}{\rho h(0)},$
the dimensionless film height as $l = \frac{2h(t)}{P},$
the dimensionless vapor flux as $j = \frac{J}{J_0},$ and
the dimensionless velocity as $\mathbf{w} = \frac{\mathbf{v}}{v_0}$.
$l = \varepsilon ( 1 - \tau)$, $d l/ d \tau = -\varepsilon$.

The dimensionless P\'{e}clet number is defined as the ratio of the rate of advection of mass by the flow to the rate of diffusion of the mass driven by the diffusion. In our case, the P\'{e}clet number may be written as  $\mathrm{Pe} = \frac{P v_0}{2 D} =\frac{a J_0 P}{2 D \rho}.$ Concentration can conveniently be measured with respect to the initial concentration.
We denote $k_d =\tilde{k}_d P/2$.

In thin-film limit ($\varepsilon = 2 h(0)/P \ll 1$), variations of the concentration and velocity along the $z$-axis is assumed to be negligible. In this case, the problem is quasi-two-di\-men\-sio\-nal, and therefore the boundary conditions on the free surface and on the substrate can be included in the governing equations as the source or sink terms. The conservation of the solute is written as
\begin{equation}\label{eq:transferDL}
  \frac{d l}{d \tau} + \varepsilon l\nabla \mathbf{w} = - \frac{ \varepsilon j}{a}
\end{equation}
or
$$\nabla \mathbf{w} = \frac{1 - j/a}{\varepsilon (1 - \tau)}$$
and its solution should satisfy the boundary condition at the walls of the hexagonal cell $\mathbf{w}_\mathbf{n} = 0.$

The conservation of the dispersed particles is written as
\begin{equation}\label{eq:transDL}
  \partial_\tau (cl) + \varepsilon l \nabla(c\mathbf{w}) = \varepsilon\mathrm{Pe}^{-1}\left( l \nabla^2 c - k_d c\right)
\end{equation}
or
$$  \partial_\tau c + \varepsilon \nabla (c \mathbf{w} ) = \mathrm{Pe}^{-1} \varepsilon \nabla^2 c + \frac{1 - k_d \mathrm{Pe}^{-1} }{ 1 - \tau } c$$
and its solution should satisfy the boundary condition
\begin{equation}\label{eq:bcdl}
\partial_\mathbf{n} c = 0.
\end{equation}

In our approach,  Eqs.~\eqref{eq:transferDL} and \eqref{eq:transDL}  are independent and may be solved separately. To solve the partial differential equations~\eqref{eq:transferDL} and \eqref{eq:transDL}, we utilized finite element methods using FlexPDE Professional 6.37.

\section{Simulation results and discussion}\label{sec:results}
\begin{figure}[htbp]
 \centering
  \includegraphics[width=0.5\textwidth]{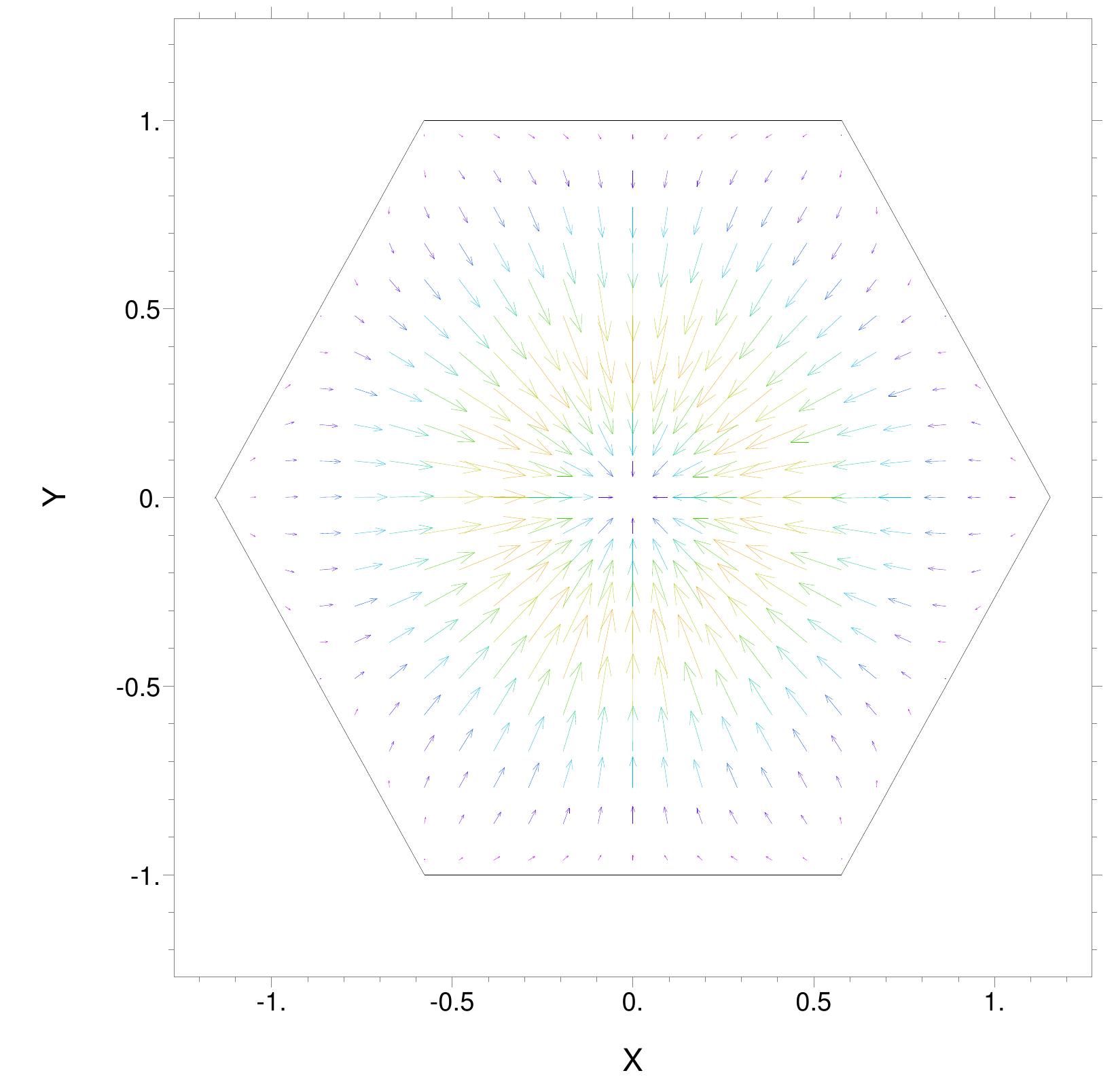}\\
  \caption{Velocity field inside a hexagonal cell ($\varepsilon = 0.01$, $P = 2 d_h$,  $\tau = 0.5$). }\label{fig:field}
\end{figure}

When a colloidal film dries under a mask, the liquid flows towards regions with the highest evaporation. Such behaviour is intuitively clear, similar to the results obtained for a droplet~\cite{Vodolazskaya2014BPFUR}, and follows the mass conservation (Fig.~\ref{fig:field}).

\begin{figure}[htbp]
 \centering
\includegraphics[width=0.5\textwidth]{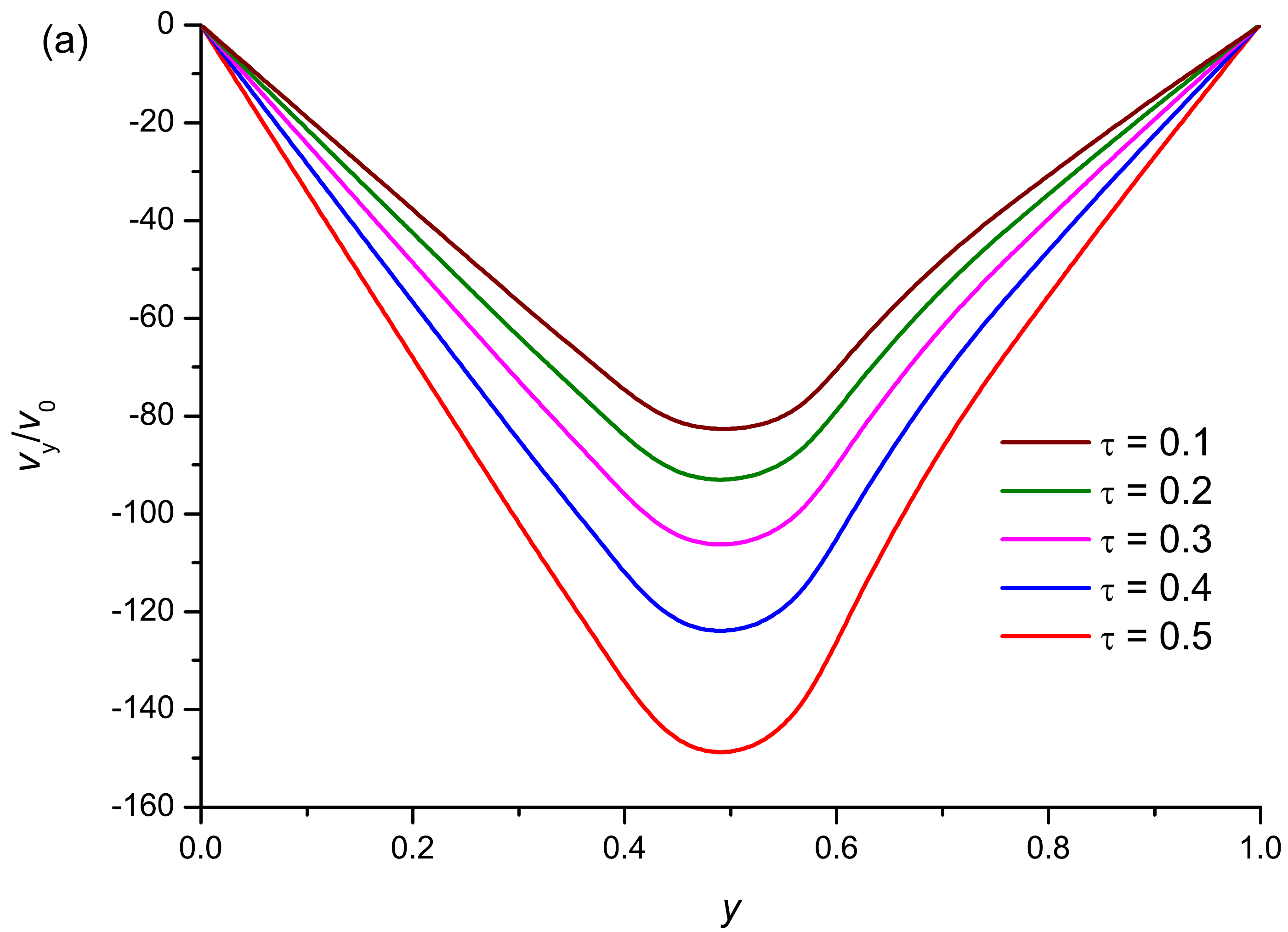}\hfill\includegraphics[width=0.5\textwidth]{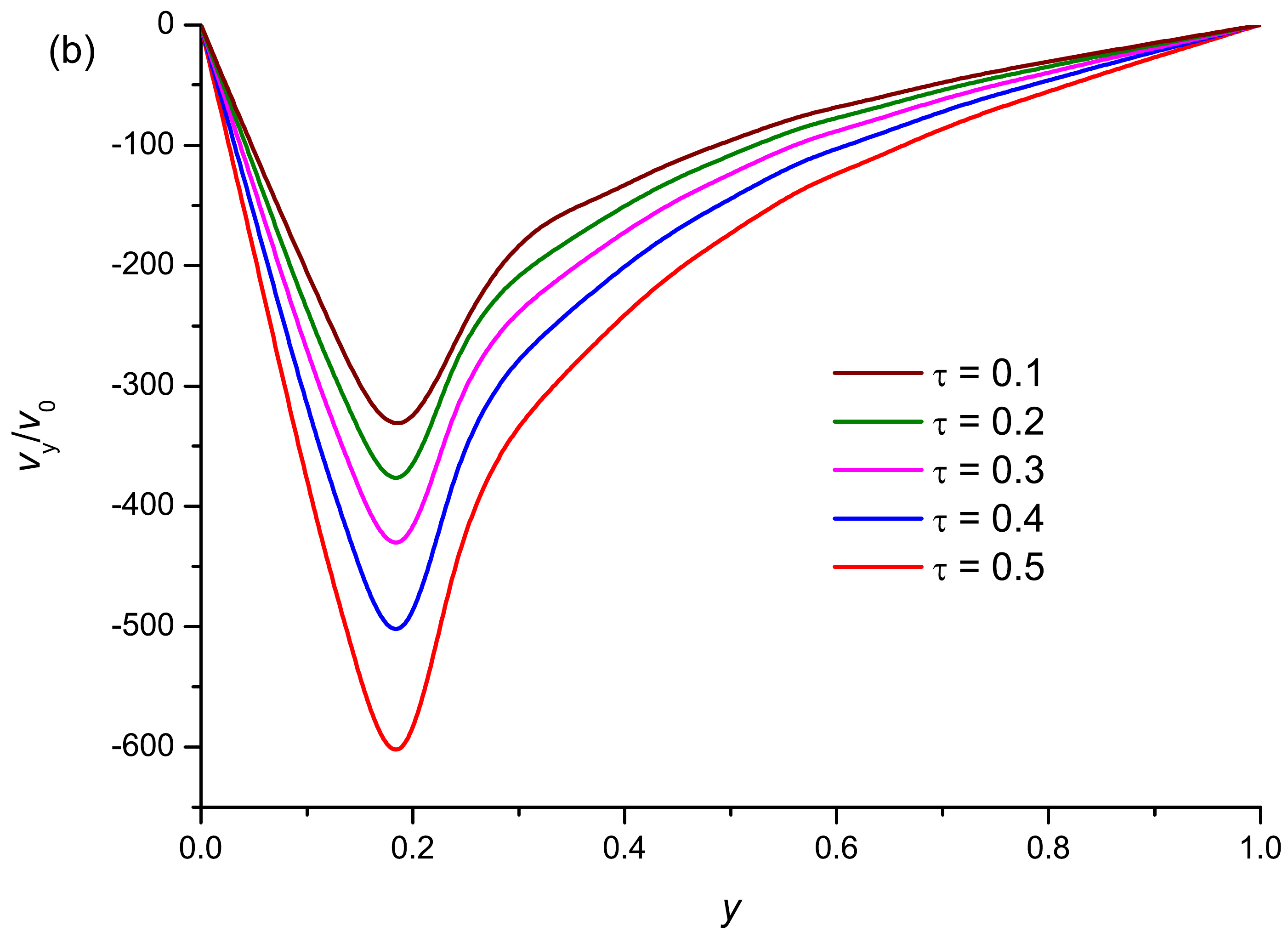}\\
  \caption{Velocity along $y$-axis for different times ($\varepsilon = 0.01$) (a) $P = 2 d_h$,  (b) $P = 6 d_h$. }\label{fig:flow}
\end{figure}

Inside a hexagonal cell the liquid flows towards the centre. The highest velocity corresponds to an edge of the hole  (Fig.~\ref{fig:flow}).

\begin{figure}[htbp]
 \centering
  \includegraphics[width=0.5\textwidth]{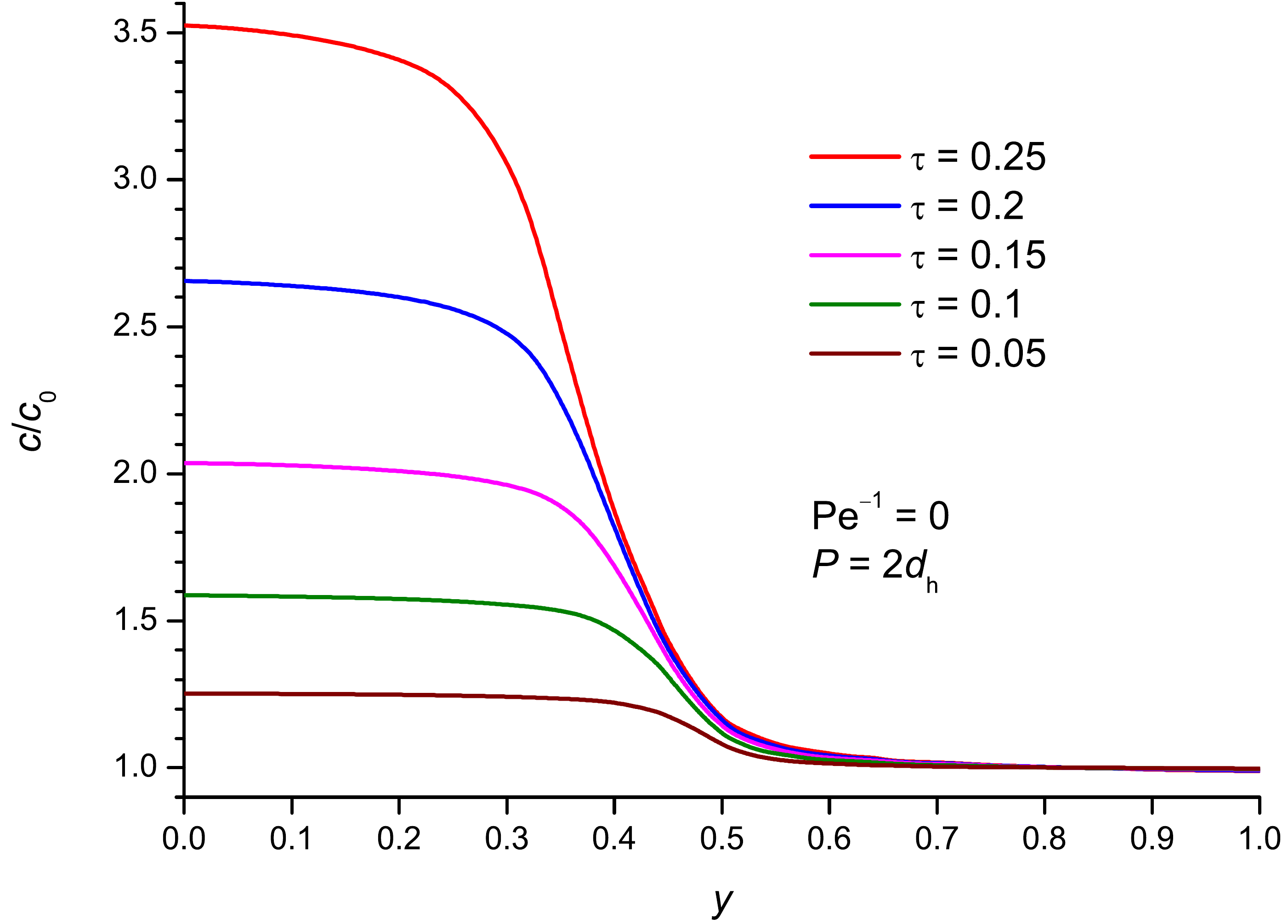}\\
  \caption{Concentration of the dispersed particles inside the thin film for different times ($\varepsilon = 0.01$, $P = 2d_h$) ($k_d = 0$).}\label{fig:advection}
\end{figure}

Having the velocity field, we can simulate the transfer of particles.
To simplify our consideration, we start with the situation when both the diffusion and sedimentation are negligible ($\mathrm{Pe}^{-1}=0$, $k_d=0$), i.e. the right hand side of Eq.~\ref{eq:transDL} is equal to zero. It is natural to expect that the particles will accumulate under the holes. Our computations demonstrate that the concentration of the particles under the holes increases by several times during half of the total time taken for evaporation (Fig.~\ref{fig:advection}).

\begin{figure}[htbp]
 \centering
  \includegraphics[width=0.5\textwidth]{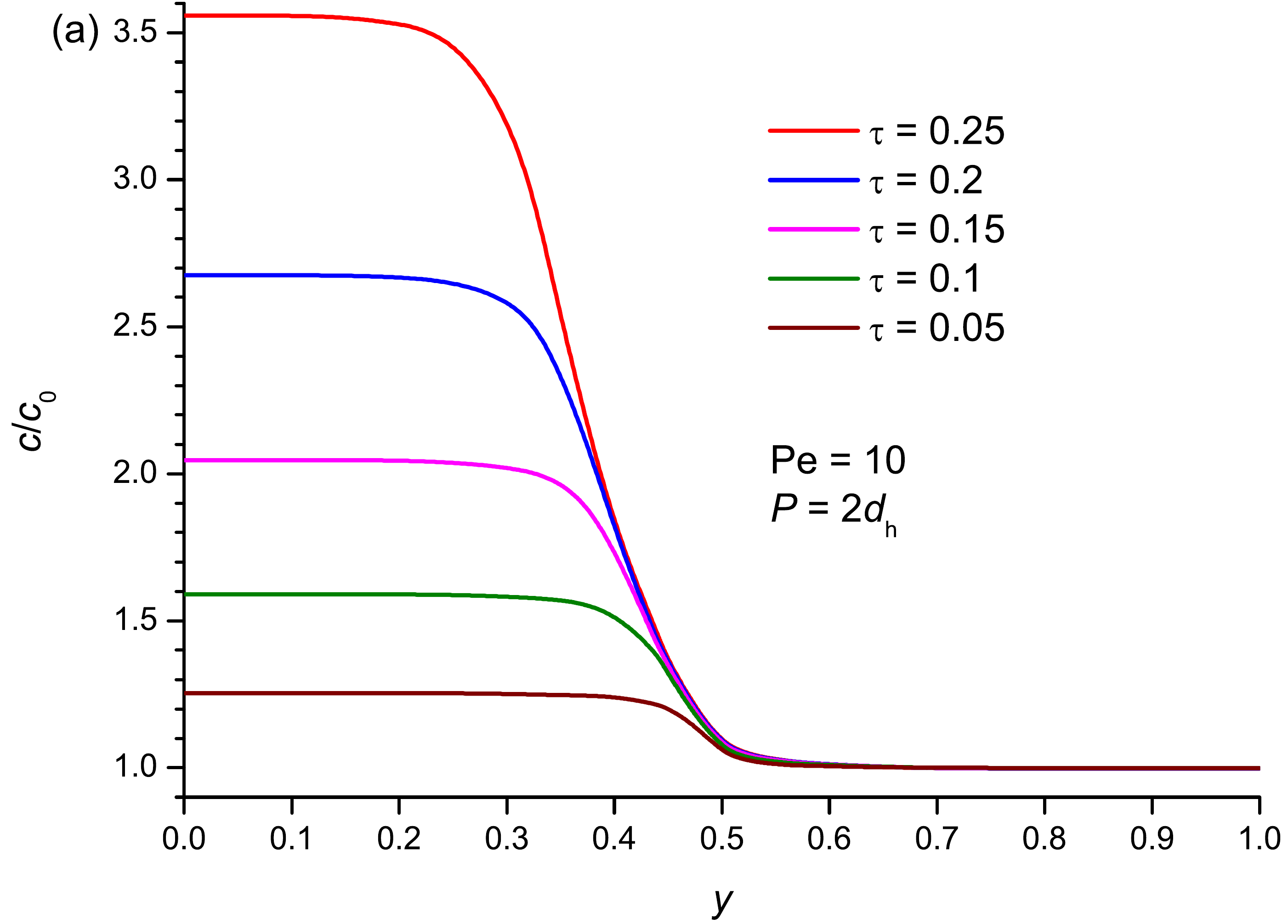}\hfill\includegraphics[width=0.5\textwidth]{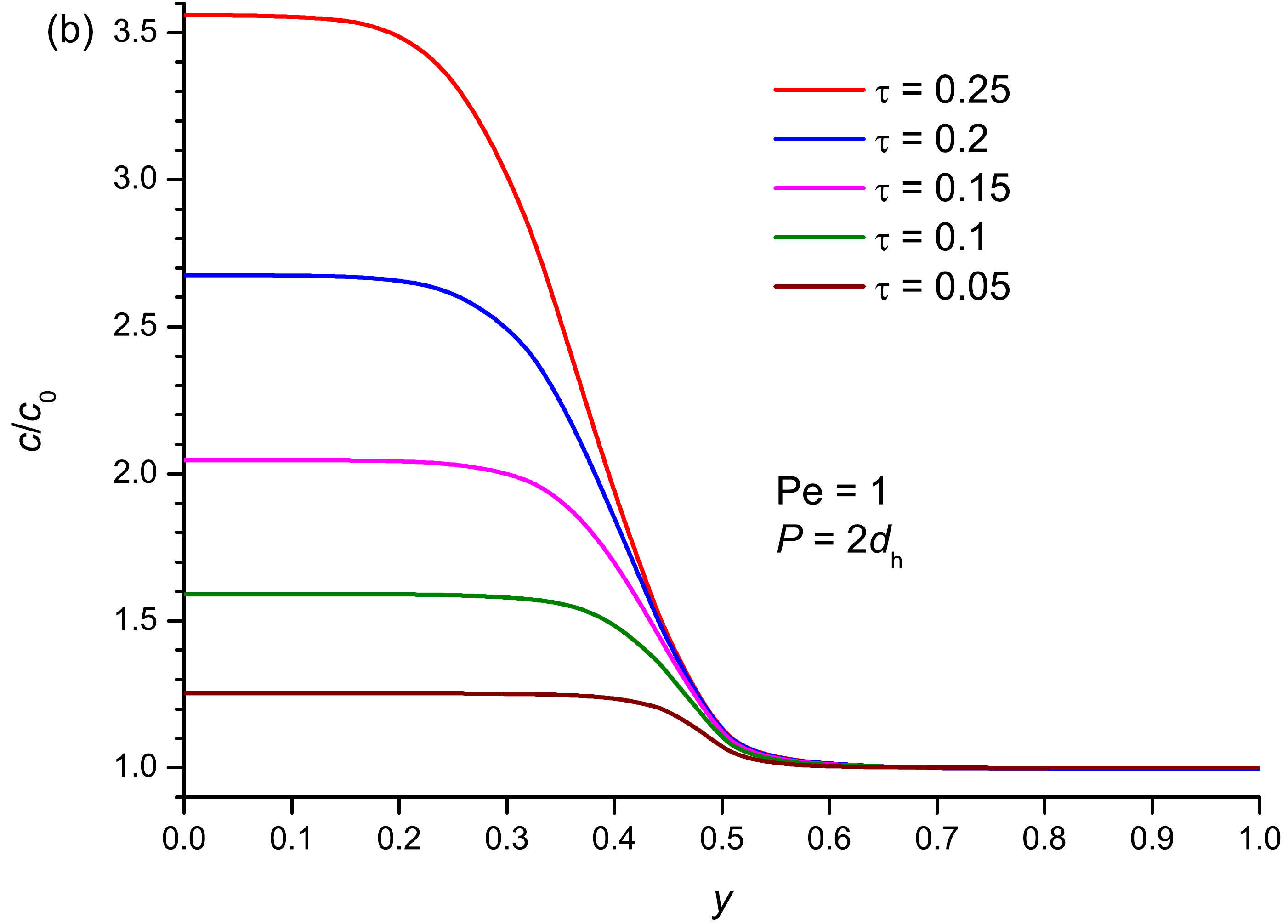}\\
  \caption{Concentration of the dispersed particles inside the thin film for different times ($\varepsilon = 0.01$, $P = 2d_h$); (a) $\mathrm{Pe}=10$,  (b) $\mathrm{Pe}=1$.  The sedimentation is absent ($k_d = 0$).}\label{fig:advectionPe}
\end{figure}

We examined the effect of diffusion for different values of the P\'{e}clet number (Fig~\ref{fig:advectionPe}). When the dispersed particles have high diffusivity, advective transfer could be hindered by diffusion. Nevertheless, this effect is visible only in a narrow region near the hole edge, where the concentration gradient is large.

\section{Conclusion}\label{sec:concl}
The proposed model allows a qualitative description of the real system. Nevertheless, it has some obvious shortcomings originating from the assumptions stated in Section~\ref{sec:model}. There are different ways to improve the model. First of all, one should take into account that the free surface is not perfectly flat. The flow inside the film is actually caused by the curvature of the surface shape. Moreover, the liquid should be considered as viscous. In this case, the velocity will vary from the bottom of the droplet to its apex, hence  the quasi-two-di\-men\-sio\-nal approach would hardly be acceptable.  The viscosity, the surface tension, and the diffusivity depend on the particle concentration. These dependencies should be taken into account. The dependence of the surface tension on concentration may produce Marangoni flow. The thermal effects ignored in the present investigation may be the cause of convective flows. The lubrication-approximation~\cite{Anderson1995} may be used as one of the possible ways to incorporate these effects into our consideration. This approach has been successfully applied to  desiccated colloidal droplets with axial symmetry~\cite{Tarasevich2013,Fischer2002,Ozawa2005}.
In further studies, the speculative evaporative flux~\eqref{eq:J} should be replaced with a calculated one.
Furthermore, the solute particles occupy finite volume. When we take into account this effect, the governing equations are no longer independent and should be solved together.
Taking account of such additional effects significantly complicates the model, but in reality leads to little  a significant change in the results.  Thus, although we can identify a number of different ways to improve the proposed model, we believe that this model is useful even in the present simplest form.

\section*{Acknowledgements}
The reported research was supported by the Ministry of Education and Science of the Russian Federation through Project No.~643.

\section*{References}
\bibliography{drops}

\end{document}